\documentclass[12pt,preprint]{aastex}

\slugcomment{Accepted: ApJ, November 2011}

\shorttitle{Cepheid PL Slope and Width }
\shortauthors{Madore \& Freedman}

\begin{document}


\title{Multiwavelength Characteristics of Period-Luminosity Relations}


\author{\bf  Barry F. Madore \& Wendy L. Freedman}
\affil{Observatories of the Carnegie Institution of Washington \\ 813
Santa Barbara St., Pasadena, CA ~~91101} \email{barry@obs.carnegiescience.edu, wendy@obs.carnegiescience.edu}



\begin{abstract}
We present a physically motivated explanation for the observed,
monotonic increase in slope, and the simultaneous (and also monotonic)
decrease in the width/scatter of the Leavitt Law (the Cepheid
Period-Luminosity (PL) relation) as one systematically moves from the
blue and visual into the near and mid-infared. We calibrate the
wavelength-dependent, surface-brightness sensitivities to temperature
using the observed slopes of PL relations from the optical through the
mid-infrared, and test the calibration by comparing the theoretical
predictions with direct observations of the wavelength dependence of
the scatter in the Large Magellanic Cloud Cepheid PL relation. In
doing so we find the slope of the Period-Radius (PR) relation is $c =$
0.724 $\pm$ 0.006. Investigating the effect of differential reddening
suggests that this value may be overestimated by as much as 10\%;
however the same slope of the PR relation fits the (very much
unreddened) Cepheids in IC~1613, albeit with lower precision. The
discussion given is general, and also applies to RR Lyrae stars, which
also show similarly increasing PL slopes and decreasing scatter with
increasing wavelength.
\end{abstract}

\keywords{Stars: variable: Cepheids  -- Stars: fundamental parameters -- Stars: variable: RR Lyrae}
.
\medskip
\medskip
\medskip
\medskip
\medskip
\medskip
\medskip
\medskip
\medskip
\medskip
\medskip
\medskip
\medskip
\medskip
\medskip
\medskip
\medskip
\medskip
\medskip
\medskip
\medskip
\medskip
\medskip
\medskip
\vfill\eject
\section{Introduction}

The Period-Luminosity (PL) relation for Classical Cepheids is a
2-dimensional projection of the higher-order Period-Luminosity-Color
(PLC) relation. The PLC relation, in turn, is a finite portion of a
plane, a so-called strip, embedded in the 3-dimensional space of
period, luminosity and color. This strip is defined and delimited by
physical limits (on temperature and luminosity) within which the atmospheres
of these stars are found to be pulsationally unstable.  A graphical
representation of how these various projections arise and are
inter-related was first given in Figure 3 of Madore \& Freedman
(1991). From this perspective it is clear that the width and
orientation of the PLC strip in 3-dimensional space will have direct
and profound influence on the measured properties of the 2D relations,
projected and multiply transformed to observable
(wavelength-dependent) quantities (magnitudes and colors).

It has long been known that the intrinsic width of the Cepheid
Period-Luminosity (PL) relation (as commonly parameterized by its
formal dispersion) {\it decreases} as a function of increasing
wavelength (see, for example, Madore \& Freedman 1991 for the optical,
McGonegal et al. 1982 for the near-infrared; and more recently Madore
et al. 2009 for applications in the mid-infared). What is less
commonly noted, although just as obvious, is the fact that the slope
of the Cepheid PL relation does the opposite, and {\it increases}
systematically with increased wavelength. This anti-correlation of
slope and dispersion is shown in Figure 1, which is an updated and
expanded version of a plot first presented as Figure 6 in Madore \&
Freedman (1991). Scatter is the quantity of most interest when any
such relation is being used for distance determinations. The slope, on
the other hand, is of little or no consequence in most practical
applications.  In the following, however, we show that the behavior of
these two quantities, slope and scatter, share a common physical
explanation, and that they can be readily understood (i.e., predicted)
from first principles. Moreover, there may indeed be some practical
implications arising from this realization.


\section{An Analytic Approach}

The total luminosity of a radiating object is the product of its total
(radiating) area times the mean surface brightness of that area.
Generalizing the Stefan-Boltzmann Law to a monochromatic (rather than
bolometric) observation can be done by parameterizing the surface
brightness by some power, $a_{\lambda}$,  of the effective temperature, $T_e$. Thus,
the entire (i.e., infinite) planar extent of the color-magnitude
diagram is mapped by

$$M_{\lambda} = -2.5~logR^2 - 2.5 \times a_{\lambda}~log T_e + b_{\lambda}~~ . . .   ~~(1) $$

\par\noindent where R is the radius of the star and $\lambda$ is the
wavelength of the observational bandpass. By imposing the following
linear ``constraints'' on the above equation one can then define the
(red/cool and blue/hot) boundaries of the instability strip: $log~R =
c~log~P + d$ and $log~T_e = -e~log~P + f$, where $P$ is the period,
and where it is to be noted that $a_{\lambda}, c$ and $e$ are all
numerically positive quantities.  Applying these external constraints
to the master equation (1) above (where $f_{red}$ and $f_{blue}$ are
respectively the red and blue intercepts of the blue and red
boundaries of the instability strip) gives

$$M_{\lambda}^{red}  = [-5~c + 2.5~e~a_{\lambda}]~log P - 2.5 \times [a_{\lambda} \times f_{red}]  + b_{\lambda} - 5~d~~  . . .  ~~(2a)$$
$$M_{\lambda}^{blue} = [-5~c + 2.5~e~a_{\lambda}]~log P - 2.5 \times [a_{\lambda} \times f_{blue}] + b_{\lambda} - 5~d~~ . . .  ~~(2b)$$

The full (wavelength-dependent) magnitude widths of the respective PL
relations, $|M_{\lambda}|$ (which are directly related and proportional
to their intrinsic dispersions) are then found to be

$$|M_{\lambda}| = 2.5 \times a_{\lambda}~|f_{blue} - f_{red}|~~ . . . ~~(3)$$

\par\noindent
Moreover, the (central) ridge-line equation of the mean PL relation becomes $(M_{\lambda}^{red} + M_{\lambda}^{blue})/2$
or

$$M_{\lambda}^{mean} = [-5~c +2.5~e~a_{\lambda}]~log P - 2.5~[a_{\lambda} \times f_{mean}] + b_{\lambda} - 5~d ]~~ . . . ~~(4)$$ 

\par\noindent
where $F_{mean} = (f_{red} + f_{blue})/2$.

\par\noindent
The slopes of the corresponding (wavelength-dependent)  PL relations are then

$$\Delta M_{\lambda}/\Delta log~P = A_{\lambda} = -5~c + 2.5~e~a_{\lambda}~~ . . . ~~(5)$$

\par\noindent The explicit dependence\footnote{For mathematical
  simplicity the above derivation implicitly assumes that the red and
  blue boundaries of the instability strip are parallel. The referee
  has pointed out that this may in fact not be the case, citing both
  theoretical (Bono et al. 2000) and observational evidence (Tammann
  et al. 2003) to the contrary. However the main point of this
  exercise is to show that $a_{\lambda}$ is common to both Equations 3
  and 5. Whether the multiplicative factor ``e'' in Equation 5 is the
  average of two different numbers (non-parallel boundaries) or
  identical numbers (parallel boundaries) does not change the main
  conclusion.} of both the slope (through Equation 5) and the
width/scatter (through Equation 3) on the same wavelength-dependent
term, $a_{\lambda}$, now becomes clear.

\par\noindent At fixed period, the total width of the PL relation at
any given wavelength is simply the difference of Equations 2a and 2b,
with the formal dispersion being a fixed multiplicative fraction of
that width.  A quick check shows that for a temperature width of the
Cepheid instability strip of 700K (or $\Delta~log~T_e = 0.05$) this
corresponds to a color width of $\Delta(B-V) = $ ~0.5~mag and a
corresponding magnitude width of $\Delta M_V = 0.8$~mag. If the stars
are uniformly distributed within and across the strip this would
correspond to a formal dispersion of $\sigma_V = 0.8/\sqrt{12} =
\pm0.23$~mag, which compares well with the observed dispersion of
$\pm0.27$~mag for the LMC V-band PL relation as given in Madore \&
Freedman (1991).

\vfill\eject
\subsection{A Graphical Approach}

Figure 2 graphically captures the substance of the above cascade of
equations, and visually illustrates just how the slopes of the PL
relations and their widths/dispersions are coupled. Three PL relations
are shown.  The lower relation (labeled 3.6 at its terminal point to
the right of the diagram) represents a long-wavelength PL relation
having the characteristic low scatter and a steep slope.  The thick
parallel lines (flanking each of the mean PL relations) represent the
upper and lower boundaries of the instability strip as projected into
the period-luminosity plane. The light, dotted lines trace the mean
relations. The thick dashed lines running diagonally across each of
the PL relations represent lines of constant color/temperature.

It is important to note that the mapping of the PL relation at one
wavelength into a PL relation as observed at another wavelength is
solely dependent on the differential response of surface brightness to
temperature across the respective wavelengths (as parameterized bf
$a_{\lambda}$ in Equation 1). Hereafter color is taken to be synonymous
with temperature. Given that the influence of temperature on surface
brightness is known to be an increasingly sensitive function of
decreasing wavelength, and given that temperature is itself a
generally decreasing function of increasing period (longer-period
Cepheids are generally cooler than their short-period counterparts),
the mapping of the 3.6$\mu$m PL relation into the B band is shown by
the two thick arrows of decreasing amplitude in going from hot
temperatures (the arrow above the circled A) to cooler temperatures
(the shorter arrow above the circled letter B). This systematic
decrease in the vector's magnitude with period (i.e., with mean
temperature) is readily seen to be responsible for the systematic
decrease in the slope of the PL relation in going from longer to
shorter wavelengths (as explicitly given in Equation 5).

The vectors originating from points in the long-wavelength PL relation
having the same temperature (as tracked by the dashed lines) have (by
definition) exactly the same magnitude/length. Those vectors are shown
terminating in the B-band PL relation just below the circled letters C
and D. In other words, the slope of the lines of constant temperature,
$\Delta mag/\Delta$ logP, are independent of the wavelength of the PL
relation.

There is a fixed temperature difference in crossing the instability
strip at a given period. This manifests itself as an increasingly
larger magnitude width of the PL relation when observed at
progressively shorter and shorter wavelengths (as can be seen in
Equation 3).

There is also, as noted above, a general decrease in temperature
(increase in color) along the PL relation as one increases the
period. This results in the decreased magnitude of the correction for
temperature as a function of period, which in turn manifests itself as
a reduction of the overall slope of the PL relation as one goes to
shorter and shorter wavelengths.

The significant point here is that both of the above effects are
controlled by a single physical parameter, the temperature.
Accordingly, an increased slope of the Cepheid PL relation must be
accompanied by a causally connected decrease in the intrinsic
dispersion of the PL relation. The longer-wavelength PL relations will
be steeper and their intrinsic scatter must be smaller than their
shorter-wavelength counterparts.

\section{A Calibration and Tests}

Looking at Equation 4 more closely it is clear that the slope of the
PL relation is made up of two linearly additive, but physically
independent parts: (1) a constant term ``5c'', which is essentially
the slope of the Period-Radius (PR) relation converted into a
Period-Area (PA) relation, and then cast into a magnitude (i.e., by
multiplying log R by 2 $\times$ -2.5), and (2) the slope of the
Period-Color relation (i.e., $-2.5 \times e \times (a_{\lambda_1}
-a_{\lambda_2})$). It is this last term that contains the
wavelength-dependent index, $a_{\lambda}$, characterizing the
sensitivity of surface brightness to temperature variations for any
given bandpass. If the (wavelength-independent) slope of the
Period-Area relation is simply subtracted from the
(wavelength-dependent) PL slopes one would be left with the purely
wavelength-dependent, Period-Surface-Brightness relation (PSB) slope
(for instance, transforming Equation 1 gives $M_{\lambda} +2.5~logR^2
= - 2.5 \times a_{\lambda}~log T_e + b_{\lambda} $). Given that the
constant of proportionality between slopes of the PSB relations and
their respective dispersions must be the same number, independent of
bandpass, we have then a means of determining the slope of the PA
relation by demanding that the ratio of the PSB slope to PL dispersion
must be constant and independent of wavelength.

Published values of the slope of the PR relation have been recently
tabulated by Molinaro et al. (2010). Their own work suggests values of
$c$ ranging from 0.71 to 0.75, with historical values, drawn both from
theory and observations, ranging more widely from 0.67 to 0.77.  We
now explore the run of the aforementioned proportionality constant
with bandpass, parameterized by the input slope of the PR relation.
The abcissa in Figure 3 which is the ratio of slope of the PL relation
(Equation 5) to the dispersion in the PL relation (effectively
Equation 3), will become independent of wavelength when the correct
value of the scaled slope of the period-radius relation (i.e., -5c) is
subtracted from the numerator, as given in Column 3 of Table 1.  The
intent is to find a slope of the PR relation that minimizes any
residual trend of the proportionality constant with
bandpass/wavelength. The circled dots in Figure 3 show that solution:
no trend with bandpass (a formal slope of 0.000 $\pm$ 0.015) for an
input value of c = 0.721, and a resulting scatter ($\pm$ 0.009) that
is exceedingly small. In changing the slope of the PR relation to
correspond to the Molinaro et al. limits, as given above, the two
flanking solutions are realized. In both cases the dispersion is much
enhanced, and there is a strong (and unphysical) trend seen with
wavelength. This already small range in the PR slope (of only $\pm$3\%
peak-to-peak) is now much more narrowly defined by the present study.
Our final solution for the slope of the Cepheid Period-Radius relation
and its error is c = 0.724 $\pm$ 0.006.

The results of the above fitting, now adopting a PR slope of 0.724,
are given in Table 1.  The first column gives the bandpass, ranging
from the blue to the mid-infrared, at which each fit was made.  In the
second column we list, in order of increasing wavelength, the
published slopes, $A_{\lambda}$ (equal to $-5c + 2.5 e a_{\lambda};$ see
equations (4) and (5)) of PL relations taken from a merger of Madore
\& Freedman (1991) for the optical, Persson et al. (2009) for the
near-infrared, and Madore et al. (2009) for the mid-infrared. Column 3
gives the slope of the PSB relation obtained by subtracting the
constant and wavelength-independent slope of the Period-Area relation
(i.e., -5 $\times$ 0.724 = -3.62) from the respective PL slopes (as
given in Column 2) leaving $A_{\lambda} + 3.62 = 2.5 e a_{\lambda}$. As
already derived above, the multi-wavelength slopes of the PSB
relations and the dispersion/widths of the PL relations must each be
proportional to the same factor, $a_{\lambda}$. As can be seen in
Figure 3 the incorrect choice of the slope of the PR relation very
quickly leads to a divergence with wavelength in the prediction of the
scale factor (that must be independent of wavelength) required to
convert the slope to a scatter. This diagnostic diagram can therefore
be used to find the wavelength-independent slope of the PR relation
that best predicts the wavelength-dependent scatter in the observed PL
relations from the optical BVRI to the near-infrared JHK bands.

As can be seen in Figure 3 for a value of c = 0.72 the
Dispersion-to-Slope Ratio is independent of wavelength and has a value
of 0.316.  Using this proportionality factor  we convert each of
the wavelength-dependent PSB slopes to their corresponding
wavelength-dependent PL-relation dispersions. These predicted
dispersions are given in Column 4 of Table 1 (e.g., for a B-band
Surface-Brightness slope of 1.19 the predicted dispersion in the
B-band PL relation is 1.19 $\times$ 0.316 = 0.376~mag). Column 5 of
Table 1 contains the observed PL relation dispersions,
self-consistently taken from the same references cited above for the
slopes of the PL relations. The correspondence of the predicted and
observed dispersions is impressive. A slight divergence occurs in the
mid-infrared where the observed dispersions are upper limits, due to
the fact that these magnitudes were derived from only two random-phase
observations and that they contain additional scatter imposed by the
back-to-front depth and tilt of the LMC, whose Cepheids were used for
these demonstrations.

An independent test using Cepheids in the Small Magellanic Cloud is
compromised by the extremely large geometric, extrinsically correlated
scatter due to the extension of the SMC along our line of sight;
however, the somewhat more distant galaxy IC~1613 can be used for this
purpose. In the paper by Antonello et al. (2006) there are 48 Cepheids
with complete BVRI photometry having periods ranging from 3 to 42
days. We have used these observations to derive the slopes and
dispersions (the latter not having been published by these authors)
for the four PL relations. The results are shown graphically in the
lower left portion of Figure 2. The IC~1613 data are shown as filled
dots enclosed by squares, each of which has an error bar derived from
our calculated uncertainty on the slope at each wavelength. The points
are plotted for an input value of the PR relation taken from the LMC
solution. To within the uncertainties there is no obvious trend of the
dispersion-to-slope ratio for the IC~1613 Cepheids using the LMC PR
relation. There is, however, a systematic offset in the ratio which
can be directly attributed to a narrower sampling of the Cepheid
instability strip by the IC~1613 Cepheids. The conclusion remains that
the slope of the PR relation is identical for these two galaxies.
Extending the test to other systems with a wider range of
metallicities will be of interest. However, until GAIA produces
pallalaxes for a sufficiently large sample of Milky Way Cepheids
calibrating this relation in the galaxy will have to be postponed
until secure reddenings and accurate distances are availbe for more
than a handful of galactic Cepheids.

\subsection{The Effects of Differential Reddening}

Differential reddening with the sample of (LMC) Cepheids used in this
study will contribute to the observed scatter in the respective PL
relations.  This extrinsic contribution will also decrease in its
magnitude as a function of increasing wavelength, in parallel with the
decreased intrinsic width with wavelength driven temperature
sensitivity. Here we explore the systematics of differential reddening
on this study.

Independent estimates of the total line-of-sight dispersion in the
reddening to the LMC (e.g., Grieve \& Madore 1986) suggests that
$\sigma_{A_V}$ is on the order of $\pm$0.10~mag.  We have
progressively removed (in quadrature) increasing amounts of possible
differential reddening bracketting this value, and iteratively
re-solved for the input slope of the PR relation using the minimum
variance condition on the derived value the scaling factor ``c''. This
was followed by visually confirming with the wavelength plots
(individually corresponding to Figure 3) that there was no residual
correlation with wavelength.  For a one-sigma dispersion in the
putative differential reddening of $\sigma_{A_V} = 0.10$~mag the
minimum dispersion on ``c'' was found to be $\pm 0.0040$ for c =
0.754.  Recall that the minimum dispersion for the solution with no
differential reddening correction was marginally smaller (at a value
of 0.0036 for c = 0.724). The $\sigma_{A_V}$ correction of 0.10~mag
results in a 4\% change in the derived value of the slope of the PR
relation. Increasing the differential reddening correction to
$\sigma_{A_V} = 0.20$~mag more than doubles the error on the fit
taking it to 0.0074 for a minimization value of c = 0.792, which is
now a 10\% systematic shift in the derived value of the slope of the
PR relation.

We conclude that correcting for a reasonable amount of differential
reddening (i.e., $\sigma_{A_V} = 0.10$ mag.) acts to systematically
increase the slope of the derived PR relation (but by only
4\% in this instance) at the expense of decreased precision in the
final fit. However, there is an alternative approach to dealing with
differential reddening.

The Wesenheit function, $W = V - R \times (B-V)$ (Madore 1976) is
designed to eliminate the effects of dust (total and differential) by
a judicious choice of scaled colors and magnitudes, adjusted to have
reddening (in the color) exactly cancel extinction (in the magnitude),
where $R = A_V/E(B-V)$. But, that same formalsm of combining colors
and magnitudes can alternatively be used to target and scale out the
temperature variation across the PL relation (at fixed period), such
that $W' = V - \beta \times (B-V)$ where $\beta$ is the slope of lines
of constant period crossing the Cepheid instability strip within the
color-magnitude diagram. In the absence of differential reddening $W'$
would have minimum dispersion when its slope is equal to the slope of
the PA relation, namely ``5c''.

For the LMC Cepheids with B and V photometry we have examined the run
of dispersion in $W'$ as a function of $\beta$. It reaches a minimum
dispersion of $\sigma_{W'} = \pm 0.13$~mag for $\beta = 2.15$, where
the slope of the $W'$-Period relation is determined to be -3.48 $\pm$
0.04. This would then correspond to a PR relation slope of $c = 0.70
\pm 0.01$. Noting that the minimized scatter in $W'$ is still larger
than the scatter expected for the PA relation alone, and larger than
the scatter seen at longer wavelengths ($\pm$ 0.08~mag), it is
probable that photometric errors, combined with residual reddening
effects, are limiting this determination. Nevertheless, the value of
the slope of the PR relation inferred from this one application is
also gratifyingly close to recent direct determinations.

\section{Discussion and Conclusions}

While it could have been predicted, it is only now that we know that
the decreased width and the increased slope of the Cepheid PL
relation, as a function of increased observing wavelength, is
inevitable. Both effects are driven by the wavelength-dependent
temperature sensitivity of stellar surface brightness.  At the longest
wavelengths the PL relation asymptotically parallels the
(wavelength-independent) PR relation; at shorter
wavelengths the PL relation is increasingly dominated by the (highly
wavelength dependent) surface-brightness sensitivity to temperature.

It is of interest to use aspects of this formalism to put constraints
on the slope of the PR relation. Since the dispersion must be
non-negative the slope of the PA relation cannot be be shallower than
the steepest (longest-wavelength) slope of the PL relation. The
steepest slope in Table 1 is -3.49 seen at 8.0$\mu$m. This would
require that the slope of the PR relation be steeper than c = -0.70,
since there must still be at least one power of $T_e$ flattening the slope
of the PL relation, even at the longest-wavelength limit. Examination
of the minimum dispersion solution of the generalized Wesenheit
function gives a value of $c = -0.70.$ And, more comprehensively, by
requiring that the run of dispersion-to-slope be independent of
wavelength gives for LMC Cepheids $c = -0.724 \pm 0.006$, a value that
also fits the Cepheids in IC~1613.

The implications of this discussion are not confined to Classical
Cepheids. For example, the somewhat surprising (at the time) discovery
by Longmore, Fernley \& Jameson (1986) that RR Lyrae stars obey a
fairly steep and tightly-defined PL relation in the near infrared, as
compared to the same stars whose optical magnitudes show almost no
dependence on period, is directly explained by the same argument given
here in the main text. Inspection of the plots given by Longmore,
Fernley \& Jameson (1986) would suggest that the scatter in the derived
absolute magnitudes for RR Lyrae stars decreases by at least a factor
of two in going from V to K as the slope of the PL relation rises from
being essentially flat at V to having a value of about -2.5 at K. This
observed trend is re-enforced and confirmed by multi-wavelength
modelling of RR Lyrae stars by Catelan, Pritzl \& Smith (2004) whose
Figure 2 dramatically tracks the systematic change of increasing slope
and decreasing scatter occurring progressively from the ultraviolet to
the near infrared.

One note of caution is necessary. As the referee of this paper has
suggested additional (magnitude) scatter may compromise these
correlations in the case of RR Lyrae stars ``caused by off-ZAHB
evolution (which depends in turn on HB type) rather than intrinsic
effects.''  The same caveat may apply to the Cepheids themselves given
that stars of a given mass may make several crossings of the
instability strip giving rise to first-, second- and third-crossing
scatter in period at a given luminosity and temperature.

Finally, we note that changes in the mapping of temperature to surface
brightness might be modulated by additional physics, such as line
blanketing changes in response to atmospheric metallicity variations,
or surface gravity, effects especially in the near ultraviolet, etc. As
such, both the intrinsic width and the slope of the Cepheid PL
relation will each be physically (and mathematically) forced to
respond in a well determined fashion.  If the slope is thought to be
increased, say, at a given wavelength because of a metallicity
difference, then the prediction is that the intrinsic dispersion will
be decreased, and vice versa. However, uncertainties in differential
reddening within any given Cepheid sample will always make these
higher-order tests challenging.

\medskip
\medskip
\centerline{\it Acknowledgements} We thank Eric Persson, Vicky
Scowcroft and Mark Seibert for providing much valued
comments on early drafts of this paper.

\vfill\eject
\centerline{\bf References \rm}
\vskip 0.1cm
\vskip 0.1cm

\par\noindent
Antonello, E., Fossati, L., Fugazza, D, Mantegazza, L., \& Gieren, W. 2006, \aa, 445, 901

\par\noindent
Bono, G., Castellani, V., \& Marconi, M. 2000, \apj, 529, 293

\par\noindent
Catelan, M., Pritzl, B.J., \& Smith, H.A. 2004, \apjs, 154, 633

\par\noindent
Grieve, G.R., \& Madore, B.F. 1986, \apjs, 62, 427

\par\noindent
Longmore, A.J., Fernley, J.A. \& Jameson, R.F. 1986, \mnras, 220, 279

\par\noindent
Madore, B.F. 1976, Royal Greenwich Observatory Bulletins, 182, 153

\par\noindent
Madore, B.F. 2009, \apj, 298, 304

\par\noindent
Madore, B.F., \& Freedman, W.L. 1991, \pasp, 103, 933

\par\noindent
Madore, B.F., Freedman, W.L., Rigby, J., Persson, S.E., Sturch, L., \&
Mager, V. 2009, \apj, 695, 988

\par\noindent
McGonegal, R., McAlary, C.W., Madore, B.F. \& McLaren, R.A. 1982,
\apjl, 257, L33

\par\noindent
Molinaro, R., Ripepi, V., Marconi, M., Bono, G., Lub, J., Pedicelli,
S., \& Pel, J. W. 2010, astro-ph:1012.4376

\par\noindent
Persson, S.E., et al. 2009, \aj, 128, 2239

\par\noindent
Tammann, G.A., Sandage, A.R., \& Reindl, B. 2003, \aa, 404, 423

\vfill\eject

\begin{figure}
\begin{center}
\includegraphics [width=12cm, angle=270] {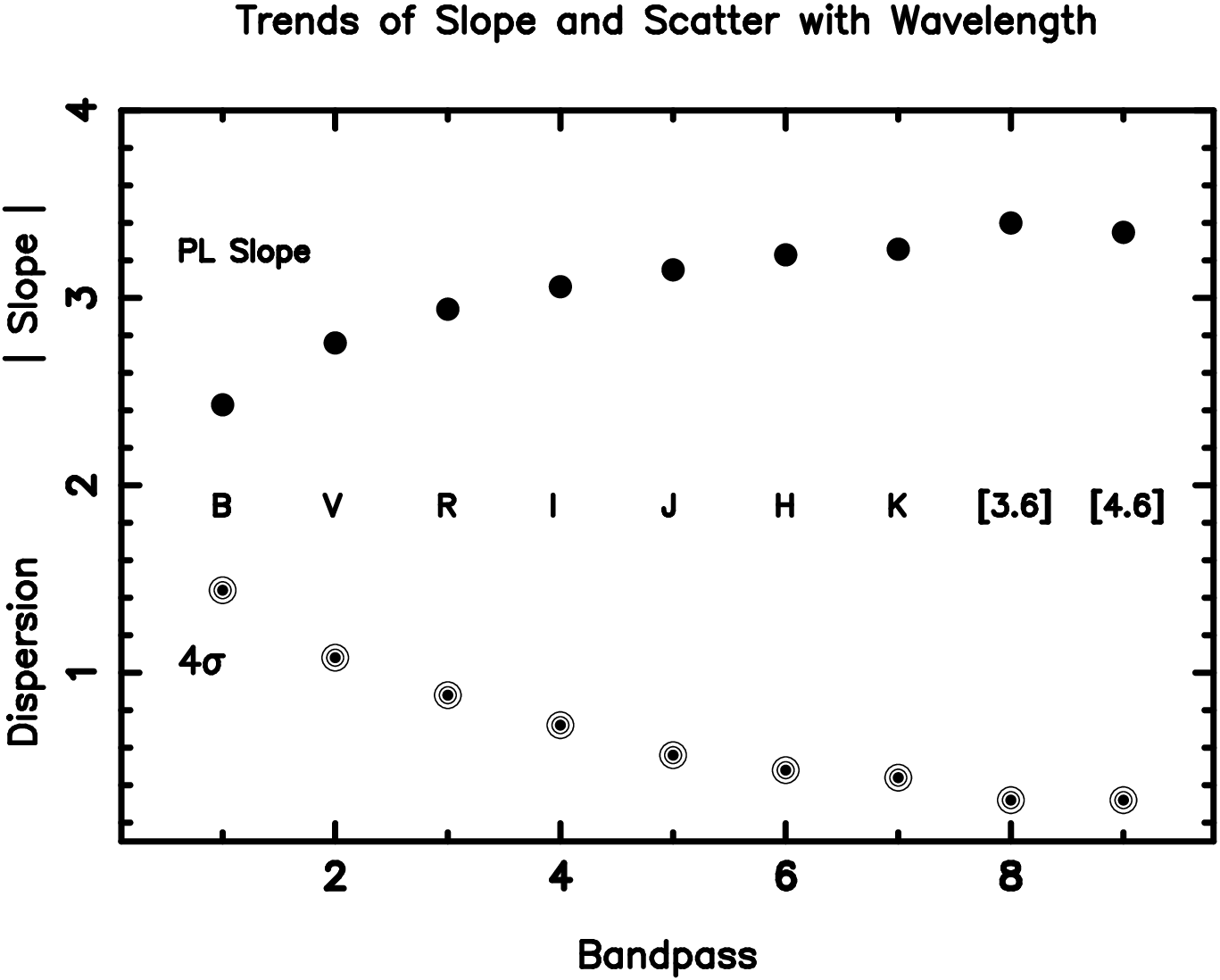}
\caption{Correlated trends in the observed slopes and widths of
multi-wavelength Period-Luminosity relations as a function of
wavelength. The upper portion of the panel shows the monotonic {\it
increase} in the slope of the Cepheid PL relation with
increasing wavelength of the bandpass. The lower portion of the figure
shows the systematic {\it decrease} of the dispersion in the PL
relation as a function of increasing wavelength.}
\end{center}
\end{figure}

\begin{figure}
\includegraphics [width=13cm, angle=270] {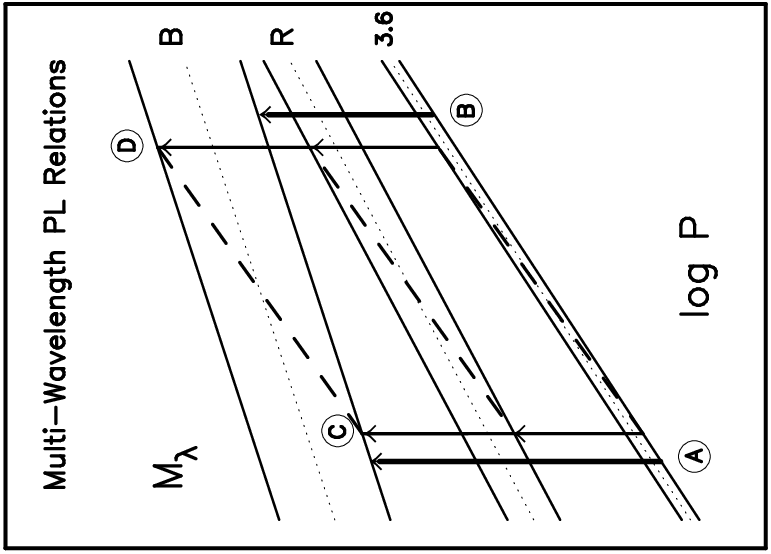}
\caption{\footnotesize Multi-Wavelength Period-Luminosity Relations. The three sets
of upward-sloping solid parallel lines represent the boundaries of the
Cepheid instability strip as projected into the period-luminosity
plane at three different wavelengths (B, R and 3.6 $\mu$m).  A line of
constant period is vertical; a line of constant luminosity is
horizontal.  Not so obvious is the fact that a line of constant
color/temperature is an upward sloping diagonal line, shown by dashed
broken lines in each of the three PL relations illustrated here.. 
The heavy vertical line above the circled letter `A', to the left
(i.e. toward shorter periods) in the diagram, represents the large
effect (at higher temperatures) of transforming the temperature
sensitivity of the surface brightness from the mid-infrared
(3.6 $\mu$m) to the B band. The shorter, heavy vertical line above the
circled letter `B', shows the decreased effect of transforming the
surface brightness at lower temperatures from the mid-IR to the B
band. This differential effect causes the B-band PL relation to have
a systematically shallower slope than that of the mid-IR PL
relation.
The heightened sensitivity of the B band to fixed temperature
differences across the instability strip at any given period results
in the significantly increased width of the B-band PL relation in
comparison to the much smaller temperature-induced magnitude width at
mid-IR wavelengths.
The width increase and the slope decrease are both controlled by the
same physics (the temperature sensitivity of the monochromatic surface
brightness); they are deterministically coupled in the sign and
magnitude of their variations.}
\end{figure}

\begin{figure}
\includegraphics [width=12cm, angle=270] {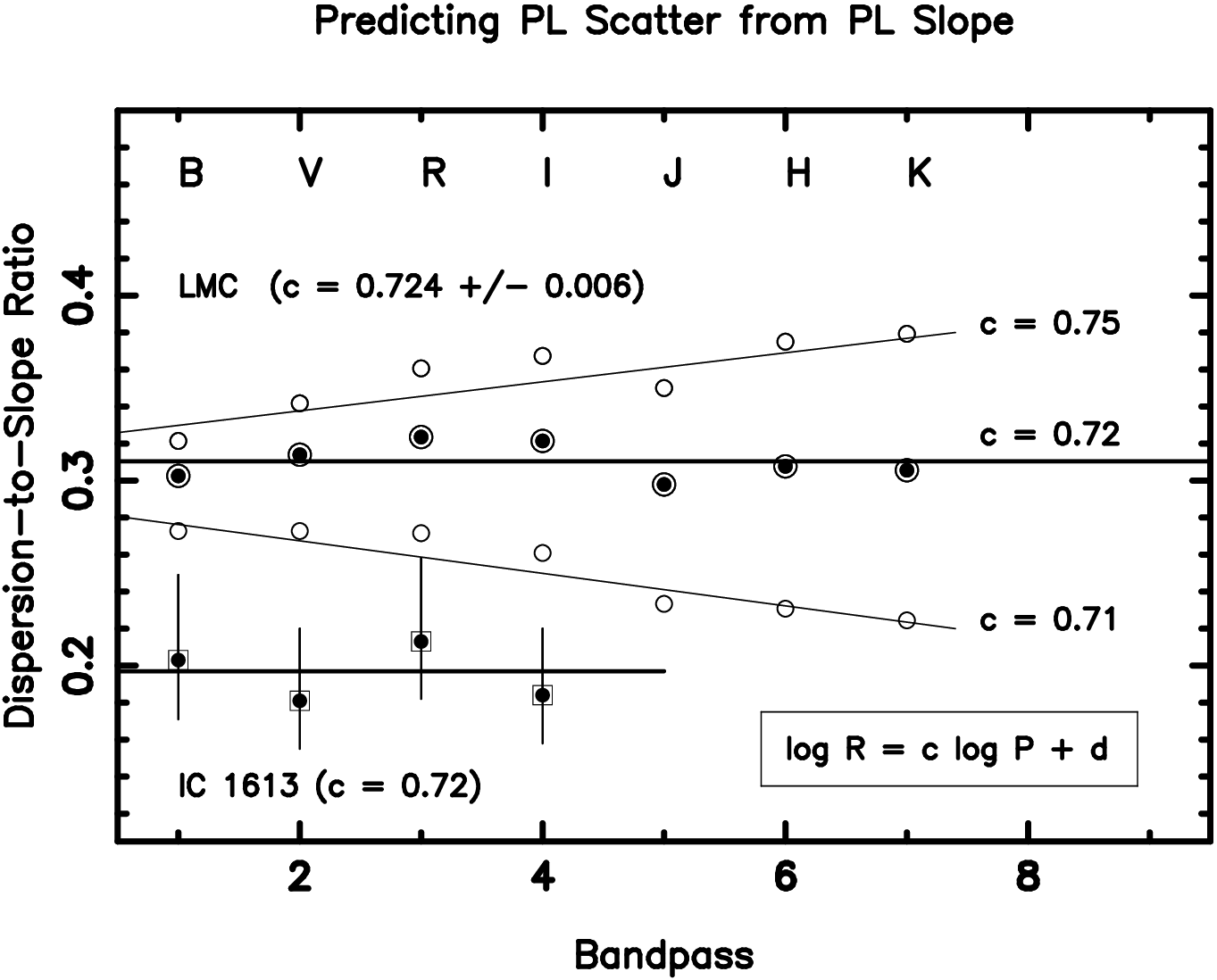}
\caption{The ratio of the dispersion in the Period-Luminosity relation
to the corresponding slope of the Period Surface-Brightness relation,
parameterized by the slope of the Period-Radius relation, ``c'', and
plotted as a function of bandpass (running from B through K as
indicated across the top of the plot). For the LMC data (top) a value
of c = 0.724 gives the wavelength-independent (flat) correlation shown
by the circled dots. The two flanking solutions (open circles)
illustrate solutions resulting from adopting the recently published
extreme values of the slope of the PR relation ($c = $ 0.71 and 0.75)
as given by Molinaro et al. (2010).  In the bottom left corner of the
plot are shown BVRI data (dot-filled squares) for Cepheids in
IC~1613. They too have a flat correlation with wavelength for the same
input value of the LMC PR relation slope; however, the offset
indicates that the width of the IC~1613 instability strip appears to
be about 30\% narrower than the LMC. }
\end{figure}


\begin{deluxetable}{ccccc}




\tablecaption{Predicted and Observed Wavelength-Dependent PL Dispersions}


\tablehead{\colhead{Band} & \colhead{PL Slope} & \colhead{SB Slope} & \colhead{$\sigma_{predicted}$} & \colhead{$\sigma_{obs.}$}\\ 
\colhead{} & \colhead{$A_{\lambda}$(obs)} & \colhead{$A_{\lambda}$(obs) + 3.62} & \colhead{(mag)} & \colhead{(mag)}\\ 
\colhead{(1)} & \colhead{(2)} & \colhead{(3)} & \colhead{(4)} & \colhead{(5)}} 

\startdata
 \\
B & -2.43 & 1.19 & 0.376 & 0.36\\
V & -2.76 & 0.86 & 0.272 & 0.27\\
R & -2.94 & 0.68 & 0.215 & 0.22\\
I & -3.06 & 0.56 & 0.177 & 0.18\\
J & -3.15 & 0.47 & 0.149 & 0.14\\
H & -3.23 & 0.39 & 0.123 & 0.12\\
K & -3.26 & 0.35 & 0.111 & 0.11\\
$[3.6]$ & -3.40 & 0.22 & 0.070 & $<$0.08\\
$[4.5]$ & -3.35 & 0.27 & 0.085 & $<$0.08\\
$[5.8]$ & -3.44 & 0.18 & 0.057 & $<$0.08\\
$[8.0]$ & -3.49 & 0.13 & 0.041 & $<$0.08\\
\enddata




\end{deluxetable}

\vfill\eject
\end{document}